\documentclass{ws-jai}
\def\btex{B{\sc IB}\TeX\ }

\usepackage[flushleft]{threeparttable}

\usepackage{graphicx}
\usepackage{subfig}
\begin{document}
\catchline{}{}{}{}{} % Publisher's Area please ignore
\markboth{WSPC}{Using World Scientific's \btex\/ Style File}

\markboth{Ligi et al.}{The operation of VEGA/CHARA\,: from the scientific idea to the final products}

\title{The operation of VEGA/CHARA\,: from the scientific idea to the final products}

\author{Roxanne Ligi$^1$, Denis Mourard$^1$, Nicolas Nardetto$^1$ and Jean-Michel Clausse$^1$}

\address{
Laboratoire Lagrange, UMR 7293 UNS-CNRS-OCA, Boulevard de l'Observatoire, B.P. 4229 F, 06304 Nice Cedex 4, France.\\
}

\maketitle

\begin{history}
\received{2013 June 29};
\revised{2013 September 3};
\accepted{2013 September 3};
\published{2014 January 6}
\end{history}

\begin{abstract}
We describe the data flow in the operation of the VEGA/CHARA instrument. After a brief summary of the main characteristics and scientific objectives of the VEGA instrument, we explain the standard procedure from the scientific idea up to the execution of the observation. Then, we describe the different steps done after the observation, from the raw data to the archives and the final products. Many tools are used and we show how the Virtual Observatory principles have been implemented for the interoperability of these softwares and data bases.
\end{abstract}


\section{Introduction}
\noindent
Optical long baseline interferometry suffers from an image of a complex observing technique reserved to specialists of instrumentation. With the advent of widely open large facilities such as VLTI and KECK, this has change a lot these recent years. Interferometric instruments are now not only used by small teams but are widely shared in open time competition. This is true also on the Center for High Angular Resolution Astronomy (CHARA) Array \citep{chara} where our instrument, VEGA (Visible spEctroGraph and polArimeter) \citep{mourard09,Mourard2011}, has been in operation for many years. This effort of opening the access to a wider community has driven some of the developments we have done these last years.

An interferometric program relies on an interferometric infrastructure and on one or many instruments. VEGA operates either with $2$, $3$ or $4$ telescopes depending on the science objectives. In all these modes, the choice of the telescopes among the $6$ that CHARA is operating opens a large number of possibilities. It's clear that, starting with a scientific idea, the problem of defining the best observing strategy is not obvious and needs tools. This is usually done by community softwares that simulate the interferometer. Moreover, doing the observations and optimizing the night with the different programs, is also a complicated task. In order to facilitate these two main aspects of the VEGA operation, we have developed a global vision of the information and data flow, starting from the scientific idea up to the final data.

\begin{figure}
\centering
	\includegraphics[scale=0.4]{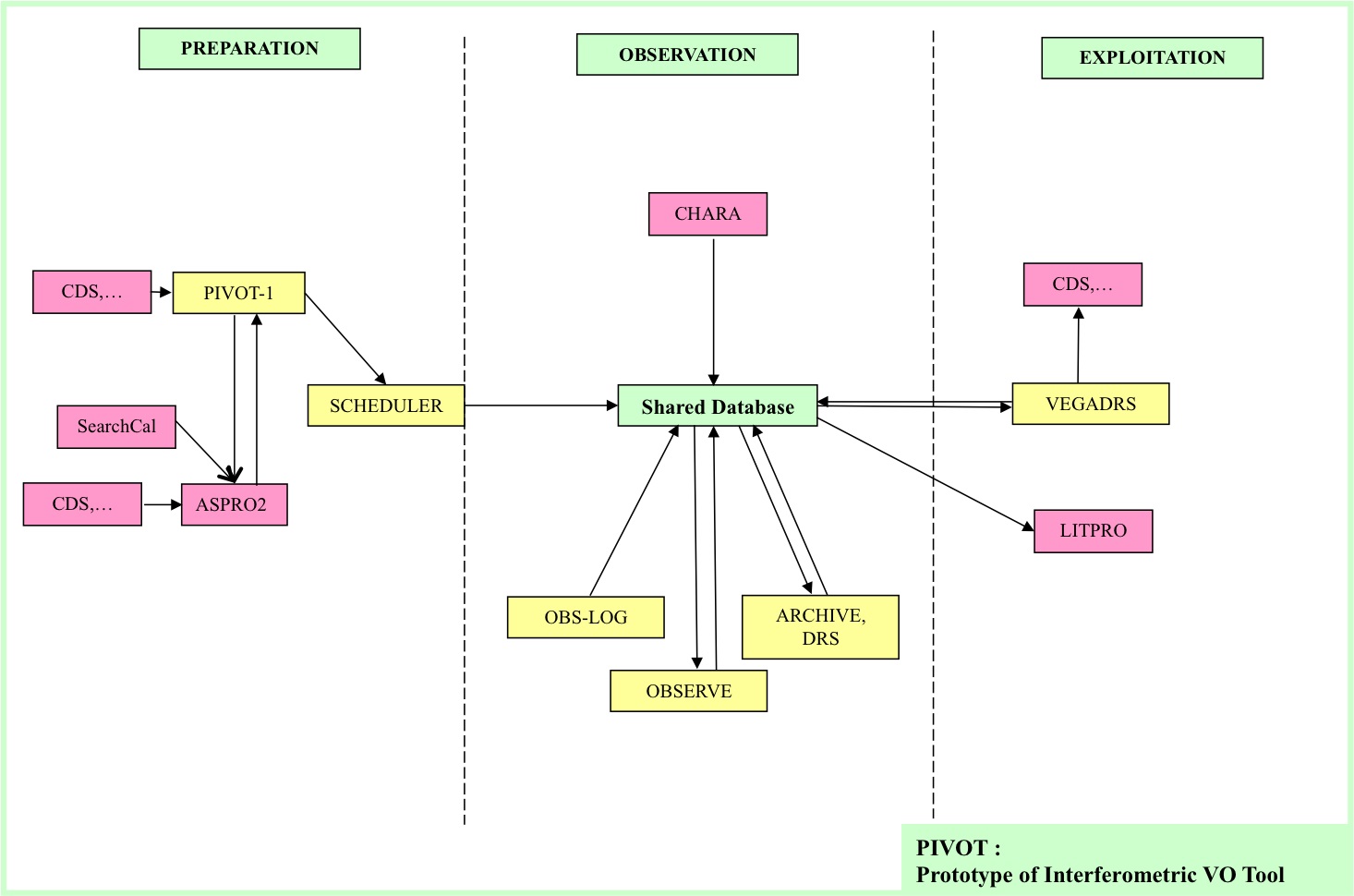}
	\caption{Diagram of the relations between all the tools used from the preparation of observations to the exploitation of data. The arrows represent the data flow between the different applications.}\label{fig:PIVOT}
\end{figure}

This paper aims to present the different steps of the VEGA use. We present the VEGA instrument and its main science programs in Section~\ref{vega}. Then Section~\ref{prepa} presents the preparation phase, Section~\ref{observation} details the way the observations are conducted and finally Section~\ref{processing} presents the necessary steps from the raw data to the final products. These steps are schematically summarized in Fig.~\ref{fig:PIVOT} where the relations between all the software tools and the databases are presented.

\section{The VEGA/CHARA instrument}
\label{vega}

The Center for High Angular resolution Astronomy (CHARA) of the Georgia State University operates an optical interferometric array located at the Mount Wilson Observatory that consists of six one-meter telescopes placed in pairs along the arms of a Y-shaped array. It yields $15$ baselines ranging from $34$ to $331$ m. Operating in the near-infrared with CLASSIC \citep{Sturmann2003}, CLIMB \citep{tenBrummelaar2012}, FLUOR \citep{Coude2012}, and MIRC \citep{Monnier2004}, and in the visible with PAVO \citep{Ireland2008} and VEGA \citep{mourard09}, the CHARA array allows a maximum angular resolution of $1.3$ and $0.3$ millisecond of arc in the K and V band, respectively.

The spectrograph is designed to sample the visible band from $450$ to $850$nm and benefits from three spectral resolutions, although the lowest one is no more used because of the difficulty of getting an accurate enough group delay tracking. It is equipped with two photon counting detectors looking at two different spectral bands. The main characteristics are summarized in Tab.~\ref{tabsp}. The simultaneous operation of the two detectors is only possible in high and medium spectral resolution. For instance, the optical design allows simultaneous recording of data (in medium spectral resolution) of the spectral region around $H_\alpha$ with the red detector and around $H_\beta$ with the blue detector. Observing in the blue requires good seeing conditions but increases by $30\%$ the limit of spatial resolution of the instrument with respect to its operation around $700$nm. The limiting magnitudes of VEGA/CHARA are presented in Tab.~\ref{perflim}. They of course highly depend on the actual seeing conditions and of the intrinsic target visibility.

\begin{table}[h]
  \centering
  \begin{tabular}{lcrrr}
  \hline
  \textbf{Grating} & \textbf{R} & \textbf{$\Delta\lambda$ (Blue)} & \textbf{$\Delta\lambda$ (Red)} & \textbf{$\lambda_R-\lambda_B$} \\
  &  & [nm] & [nm]  & [nm]  \\
  \hline
  R1\,: $1800$gr/mm & $30000$ & $5$ & $8$  & $25$ \\
  R2\,: $300$gr/mm & $6000$ & $30$ & $45$  & $170$ \\
  \hline
  \end{tabular}
 \caption{Spectral resolution (R) and bandwidth ($\Delta\lambda$) of the VEGA spectrograph, as well as the spectral separation between the two detectors.}
  \label{tabsp}
  \end{table}

\begin{table}[h]
  \centering
  \begin{tabular}{|c|c|c|c|}
  \hline
   Resolution & R & Typical Lim. Magnitude & Best perf.\\
  \hline
   Medium & 6000 & 6.5 & 8.0\\
   High & 30000 & 4.2 & 5.5\\
  \hline
  \end{tabular}
\caption {Estimation of typical limiting magnitude as a function of the different spectral resolution modes. These values are presented for the median value of the Fried parameter r0 at Mount Wilson i.e. 8~cm. The Fried parameter represents the typical size of the coherence cells of the atmospheric turbulence. We also indicate the best performances assuming an r0 of 15~cm. }
 \label{perflim}
\end{table}

The medium ($R\simeq6000$) and high ($R\simeq30000$) spectral resolutions are well suited to perform a wavelength analysis of the interferometric signal (visibility and phase), providing a kinematical resolution of $60$ and $10$~km~s$^{-1}$, respectively. It allows to probe efficiently the radiative winds and fast rotating photospheres of hot Be stars (\cite{delaa,meilland}), the complex structure of interactive binaries \cite{bonneau}, the disk of young stellar objects \cite{abaur} and for instance the the chromosphere of K giants \cite{berio}. The medium resolution is also well suited for absolute visibility studies in order to determine very precise angular diameters or to determine the orbit of detached binaries or multiple systems. Recent results of such programs concern the study of roAp stars \cite{gamequ} or CoRoT targets \cite{bigot} and more recently exoplanet hosts stars \cite{ligi}.

Another interesting possibility is the presence of a polarimeter that could be inserted into the beam. This gives new insight into many physical processes. Many science sources are linearly polarized, in particular at a small angular scale, and the interferometric polarized signal is a powerful probe of circumstellar scattering environments that contain ionized gas or dust \cite{elias, ireland, chesneau} and of magnetic properties \cite{spin2000, spin2004}.

\section{Preparation of the observation}
\label{prepa}
In this paper we do not address the important question of the best choice of the interferometric baseline and observing strategy for a dedicated science program. This has been already described in different summer schools papers or textbooks (see for example \cite{labeyrie} and references therein). This is usually based on online software simulating the answer of an interferometer to a specific object, described either by a toy model (uniform disk, gaussian disk, elongated disk, any combination of geometric bricks...) or by fits images of brightness distribution. These softwares estimates the 2D fourier transform of the brightness distribution at the spatial frequencies sampled by the interferometric configuration (baselines, wavelength, target coordinates, time, instrumental configuration). This simulated data could then be used to estimate if the parameters of the initial model or if an image of the object could be well constrained by the specific configuration. With this in hands, an observing idea could be turned into a serie of instrumental configurations.

The VEGA instrument, as any other beam combiner, requires a careful preparation. Whatever the mode of observation, on site or remotely, the procedure is basically the same and first based on a specific dedicated software, called PIVOT\footnote{\texttt{http://pivot-ws.oca.eu/}} (Prototype of Interferometric VO Tool) developed by Observatoire de la C\^ote d'Azur (OCA). The VEGA instrument has indeed a multi-program strategy of observation. Every night is optimized to perform the maximum of observations (of different targets), but with the minimum of configuration changes. In this context, the PIVOT software is extremely useful to gather and manage all observing programs, with respect to their own priority targets and configurations.  PIVOT uses the Virtual Observatory technique and several databases, like Axis2, Java, SAMP, XML and MySql~\citep{Mourard2012}.

The ultimate aim of the PIVOT software is to help the PI of an observing program to prepare an \texttt{ASCII} file (called a starlist) containing all the information needed (for VEGA and CHARA settings) to perform a certain program \cite{Mourard2011}. It contains all the observing blocks including the spectral calibration ones. Different classes of users have been defined, and the access is specific for each one of these classes\,: the PI of the VEGA instrument, the PI of run (which means the person in charge of the organization of a series of observing nights with VEGA), the PI of night (i.e. the observer), and the astronomer (PI of the scientific program). The procedure is the following.

First, the PI of an observing program fills one or more PIVOT proposals, each consisting in as many entries as science target or different instrumental configurations. It means, he has to enter (1) the names of the stars to be observed,  (2) the telescopes and baselines to be considered, (3) the duration of the observation, and (4) the spectral configuration. Several tools can be used for that (see below). PIVOT gets the stellar fundamental parameters (HD number, magnitudes, coordinates...) in the CDS-ADS\footnote{\texttt{http://cdsweb.u-strasbg.fr/}} database, which hosts the SIMBAD and Vizier tools. They contain a very large and complete database of astronomical objects and  astronomical catalogues and tables, respectively. As already said, one PIVOT entry corresponds to a star with a given configuration.

\begin{figure}
\centering
\resizebox{0.7\hsize}{!}{\includegraphics[clip=true]{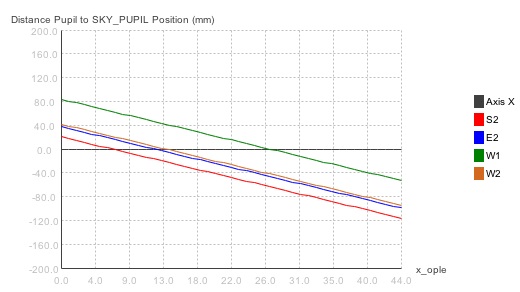}}
	\caption{The horizontal dark line represents the ideal situation. The four color lines give, for each of the four telescopes (S2, E2, W1, W2), the difference of position to the ideal situation of the reimaged pupils as a function of the position of the delay line on its stroke.}
\label{fig:pup}
\end{figure}

When all proposals are entered, the PI of the VEGA run has access to all the different configuration requests and make a classification and sorting of all the PIVOT entries. The PI takes into account the ranking from the Time Allocation Committee of CHARA and the internal VEGA classification of priorities. Using this information and the initial preparations made by the different programs, the PI is able to identify a small number of configurations able to accommodate all the priorities. He could also ask to some of the programs to change their initial request of baseline in order to simplify the global organization of the runs. For each configuration, he defines a template, associating each requested telescope to a POP and a beam. This is done taking into consideration the position of the CHARA pupils as seen by the VEGA instrument. Indeed, PIVOT provides for each telescope the predicted position of the pupil image according to the position of the cart of the delay line\,: the position of the pupil has to be the closest to zero which means that the distance between the pupil and the sky pupil position is minimized. If, for a given night, too many different configurations are required, a choice has to be made among them to reduce the number of configurations per night in order to save time. Fig.~\ref{fig:pup} presents a typical plot for a 4 telescope configuration.

The choice of POPs (Part Of Path, i.e. static additional optical paths) and beams is highly connected to the observability of all the targets to be observed in a given night. To check the observability of the stars, which means that the star elevation is higher than $30^\circ$ and that the optical paths could be equalized within the strokes of the delay lines, PIVOT and Aspro$2$\footnote{\texttt{http://www.jmmc.fr/aspro page.htm}}  exchanges data through inter operability protocols. Aspro$2$ is a tool developed by JMMC to estimate the observability of stars and to calculate visibility map of associated toy models. There is an important choice of instruments and facilities available\,: VEGA/CHARA, MIRC/CHARA, PIONIER/VLTI, AMBER/VLTI, PAVO/SUSI, ... For each instrument, all available configurations of telescopes are taken into account. Moreover, the stars parameters are extracted with an interoperability mode between Aspro$2$ and SIMBAD. Such interoperability modes (between SIMBAD, Aspro$2$ and PIVOT) are based on the SAMP messaging protocol which uses VOTable format (XML) file, both being Virtual Observatory (VO) standards provided by the Virtual Observatory Alliance (IVOA\footnote{\texttt{http://ivoa.net/}}).

The PI of run is supposed to check the observability in its globality (i.e. for all stars together). Indeed, in principle, the POPs can be changed in a given night of observation (not the beams),  but the PI of run try to reduce such changes (which require $10$ min of time) as much as possible to keep the same POP configuration, but of course, it can reduce the observability of certain targets. A compromise has to be found and the web-based tools help to take decisions on this.

Once the proposals are approved by the PI of run, the PI of program has to create the starlist file with the help again of Aspro$2$. This procedure is very easy thanks to the interoperability of several tools. First, each proposal is sent (individually) to Aspro$2$ via the interoperability mode. Then, the information are sent by Aspro$2$ to SearchCal\footnote{\texttt{http://www.jmmc.fr/searchcal}} \citep{Bonneau2006}. SearchCal, also developed at JMMC, collects all the calibrators from different catalogues, and sort them out according to the given limitations (e.g. in magnitude, coordinates, diameter...). The best calibrators found can then be sent again to Aspro$2$ to check their observability (as the calibrator is closed to the target in sky in principle, its observability is generally good, but this has to be verified) and linked them to the target. Finally, the whole set (target+calibrators) is sent back to PIVOT. An edition tool inserted into PIVOT permits correction and completion of the starlist file, for example for the number of observing blocks requested (duration of observation) and modify other parameters (recording with CLIMB, wavelength, spectral resolution, etc.). Once done, the starlist file is generated and is associated to the corresponding proposal. It also can be stored as an ascii file.

Finally, the PI of night (i.e. the observer) can prepare if necessary the strategy of observation all over the night using a specific panel in PIVOT. This step is required when the strategy over the night is particularly complex (observation of different stars at different angular hours, etc), selection of calibrators... For this purpose, the target and calibrators are selected (for each program), and the observing time is plotted as lines for each different star, the length of the segments being proportional to the duration of the observation. The strategy of observation is thus more visible and it becomes easier to organize the set of stars to observe during each night.

\section{Execution of the observation}
\label{observation}

Before the VEGA observations, the PI of the run is supposed to ($1$) define the teams of observers for all the VEGA run (in general two persons for three or four consecutive nights), ($2$) send an e-mail to the CHARA team ($48$h before the beginning of observations) specifying the configurations used (telescopes, POPs, beams, and the need or not of using CLIMB as a fringe tracker), ($3$) verify the starlist files (using a dedicated script) and put them in a specific folder on the VEGA control system, ($4$) provide to the observers the list of targets to be observed with the corresponding strategies of observations (the number and the sequence of measurements) together with the priorities in terms of targets and configurations (which depends also on the quality of the night). The preparation of the VEGA instrument itself is however done by the PI of VEGA or by a person in charge of these technical aspects (chiller, cooling of camera, etc.).

During the observations different tasks have to be done: instrument checking and setting, recording, coordination with the CHARA operator, log files, control of observing conditions (weather, humidity, observability, ...), decision on programs...

The VEGA observer has remotely access to four VEGA computers located in the electronic cabinet situated on site\,: the central computer controlling all the VEGA motors, the computer for the tracking of fringes, and the computers of the `blue' and `red' cameras, respectively. The connection to these computers has to be checked (in particular in remote) and several applications have to be launched at the beginning of the night. Operating the instrument is done using different Graphical User Interfaces talking through the internet to the different low level servers of VEGA of CHARA.

The sequence of observations is the following. The observer chooses a starlist file and inside this starlist file one of the observing block. During the slewing of the telescopes and delay lines, the setting of the VEGA instrument is done (wavelength of reference, spectral resolution, camera used, etc.). This only requires a few minutes. As soon as the first star light is locked by the CHARA tip-tilt, the VEGA instrument performs fine tuning of the alignments by controlling the position of the pupils at the entrance of the spectrograph and the correct centering of the star images on the spectrograph slit. These optimizations are done for each beam by moving remotely the corresponding M$10$ mirrors and the tip/tilt feeding mirror.

An unique capability of the CHARA array is to allow simultaneous observations with two different combiners. We thus generally use CLIMB as an external group delay tracker. As it is operating in the infrared, the visibility contrast is in principle larger than in the optical which allows to find and track fringes more easily. When CLIMB is aligned and has found the fringes offsets, a delicate operation begins in which VEGA has to be cophased with CLIMB. This is done by moving the CLIMB feeding mirrors and compensating these internal delays by moving oppositely the delay lines. In average, $30$ minutes are necessary to perform this step at the beginning of the night or after each triplet change. When cophased, VEGA can record the data.

During the recording, one has to fill the log of observation with several information, such as the PI of the program, the observed targets (and calibrators), the time of recording (beginning and end), and any specific quick comments that could help the data reduction process (quality of the fringes, loss of the star light, seeing, etc.). At the end of a sequence of observations (calibrator-target-calibrator) a spectral calibration is required which corresponds to a specific entry in the starlist file.

At the end of the night, the recorded data are archived and pre-processed automatically: several backups and also a preliminary data reduction are done. For that, the observers have to verify the consistency of their data (several folders and files are indeed generated) and send a script. The VEGA database is also automatically filled with metadata information. The database information and the logfile of the different nights permit to check the overall progress of the execution of the observations, either for an individual program in terms of number of measurement's points or for the correct execution of the top priority programs. Again at this level the decisions are taken with respect to the global ranking of the programs that the Time Allocation Committee of CHARA and that the team of VEGA co-Investigators have made initially.

\section{From the data to the final products}
\label{processing}

The OCA has developed a data reduction software (vegadrs) to process the data obtained with VEGA. It is installed both at Mount Wilson and Nice. The VEGA data processing consists of several steps. First, the raw data is converted and a spectral calibration is done. An initial processing is done to check the data quality and eventually reject short sequences of bad data. During this step, one can choose the detector (blue or red) to be considered, and one can also visualize the corresponding spectrum to check for instance the presence of spectral lines in absorption or in emission and thus decide of the processing strategy. The continuum spectrum is flattened and calibrated (polynomial laws), and the width of the spectral band (for the data reduction) is defined. Several scenarii of processing are available. The autocorrelation processes the spectral density over one spectral band, whereas the cross correlation performs a cross-spectral density between two spectral bands. In both cases, the processing is done and a sum is performed over the individual files or blocks (each block contains in principle $2500$ individual frames of 10ms). In the so-called temporal mode, the processing is performed on each block not as a whole but on a series of a certain number of frames. This mode is recommended only if the data are of good quality.

\begin{figure}
\centering
\resizebox{0.7\hsize}{!}{\includegraphics[clip=true]{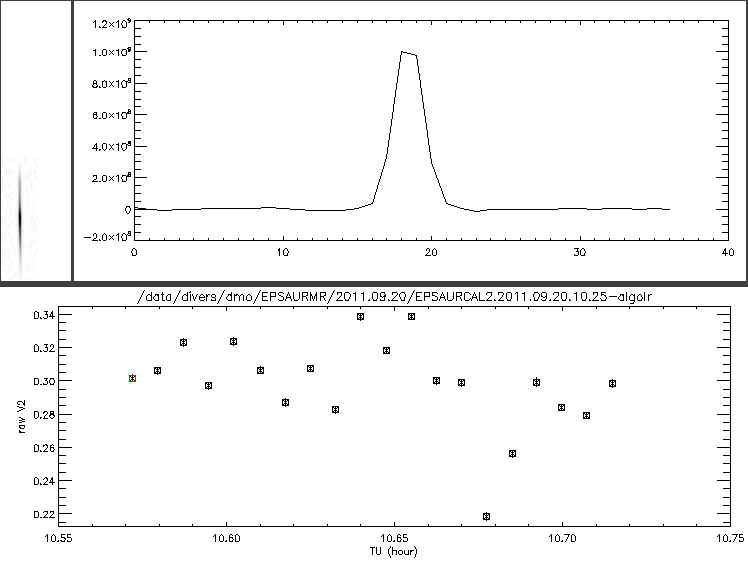}}
	\caption{A typical plot during the raw visibility analysis. The bottom panel displays the raw squared visibility as a function of time for a sequence of about 10 minutes. Each data point is an estimation of the squared visibility obtained on the serie of 2500 individual frames each being 10ms long. In the upper panel, the plot corresponds to the profile of the considered fringe peak whereas the small image on the left displays the 2D spectral density of the images around the fringe frequencies. }
\label{fig:fringe}
\end{figure}

After this, a configuration file is store and the whole processing sequence is generated as a shell script. After the processing itself, one can visualize the results with a graphical user interface that shows the fringe signal for each block and the corresponding raw visibility, the optical path difference (OPD) and the Signal to Noise ratio (S/N). Fig.~\ref{fig:fringe} presents a typical plot of an observing sequence.

It is then possible to remove the blocks of bad quality if one considers that they are damaged by instrumental or seeing conditions. When done, the results are finally ready to be analyzed. Using different \texttt{IDL} programs, raw and calibrated visibilities are estimated taking into account the selected calibrators. The final output is an \texttt{oifits} file containing the calibrated visibilities of the science star that can be read for example by the LITpro software. These oifits files are stored in the central VEGA database.

The LITpro\footnote{\texttt{http://www.jmmc.fr/litpro}} (for \textit{Lyon Interferometric Tool prototype}) software \citep{Tallon2008} has been developed by JMMC and aims at fitting models on data obtained from optical interferometers. It assumes that the direct model of the data to be fitted is known. Thus, many models are available, like uniform disk, linear limb darkened disk, ring, flatten ring, gaussian...that are all combinable together to build more complex shapes.
The user loads the \texttt{oifits} files on the software, and chooses the model and the type of data to be fitted (squared visibility, modulus of visibility, phase, Spectral Energy Distribution...). According to the selected model, various free parameters can be fixed. The output corresponds to the free parameters, along with the reduced $\chi^2$, standard deviations, covariance and correlation matrices. The software also provides graphical tools to visualize data, models, residuals, cuts in the $\chi^2$ space, etc.

The JMMC has also developed an \texttt{oifits} database in order to share the files made by the interferometric community. This tool is useful as is allows to use real or simulated data to perform modeling. Finally, after publications, some editors are requesting to publicly share the fits files associated with the observations. For some of our data published in A\&A, this is already done through the CDS.

\section{Conclusion}
In this paper we have described the sequence of actions one has to do to proceed from an interesting scientific idea using high angular resolution up to the resolution of the question with data and comparison with models. Due to the intrinsic complexity of the interferometric measurements, mainly due to the possible large number of observing modes, it has been of utmost importance, for non specialists, to highly simplify the procedure. This was exactly the aim of the work we have developed in the last years and the fact that, now, non-interferometric astronomers are already using VEGA for their scientific programs is certainly a great step forward.
Thanks to the development of interoperability by the JMMC and OCA teams, in the framework of the development of the Virtual Observatory principles, it has been possible to easily link the different applications and to use the powerful capabilities of databases to smooth all the interferometric process.

\bibliographystyle{ws-jai}
\bibliography{vegabib}

\end{document}